\documentclass[10pt,journal,compsoc]{IEEEtran}
\ifCLASSOPTIONcompsoc
  \usepackage[nocompress]{cite}
\else
  \usepackage{cite}
\fi
\ifCLASSINFOpdf
	\usepackage[pdftex]{graphicx}
	\graphicspath{{../pdf/}{../jpeg/}}
	\DeclareGraphicsExtensions{.pdf,.jpeg,.png}
\else
	\usepackage[dvips]{graphicx}
	\graphicspath{{../eps/}}
	\DeclareGraphicsExtensions{.eps}
\fi
\usepackage{amsmath}
\usepackage{url}

\usepackage{booktabs}
\usepackage{amssymb}
\usepackage{subfigure} 
\usepackage{multirow}
\usepackage{amsmath}

\makeatletter
\newif\if@restonecol
\makeatother

\usepackage[linesnumbered,ruled,vlined]{algorithm2e}
\usepackage{algpseudocode}
\usepackage{stfloats}
\usepackage{xcolor}
\newcommand{\qsl}[1]{#1}

\begin{document}
%
\title{CmnRec: Sequential Recommendations with Chunk-accelerated  Memory Network}
%
%
%
%

\makeatother
\author{Shilin~Qu*, Fajie~Yuan{*}$\dagger$, Guibing~Guo$\dagger$,  Liguang~Zhang, ~and~Wei Wei
\IEEEcompsocitemizethanks{
\IEEEcompsocthanksitem * Equal contribution.
\IEEEcompsocthanksitem $\dagger$ Corresponding authors.
\IEEEcompsocthanksitem Shilin Qu is with Monash University, Melbourne , Australia (affiliation). A part of this work was finished when Shilin was a master student at Northeastern University, an intern at Tencent Kandian Group, and an research assistant at Westlake University.
\IEEEcompsocthanksitem Fajie Yuan is with Westlake University, Hangzhou 310024, P.R. China. A part of this work was finished when Fajie was AI researcher at Tencent Kandian Group. E-mail: yuanfajie@westlake.edu.cn
\IEEEcompsocthanksitem G. Guo is with Northeastern University, Shenyang 110819, P.R. China. E-mail: guogb@swc.neu.edu.cn
\IEEEcompsocthanksitem L. Zhang  is with Kandian, PCG, Tencent, Shenzhen 518055, P.R. China. E-mail: liguangzhang@tencent.com
\IEEEcompsocthanksitem W. Wei is with Huazhong University of Science and Technology, Wuhan 430074, P.R. China. E-mail: weiw@hust.edu.cn
}
\thanks{Manuscript received April 6, 2021; revised November 22, 2021;  accepted December 23, 2021.}}

%

%

\markboth{IEEE Transactions on Knowledge and Data Engineering,~Vol.~0, No.~0, December~2021}%
{Shell \MakeLowercase{\textit{et al.}}: Bare Demo of IEEEtran.cls for Computer Society Journals}
%



\IEEEtitleabstractindextext{%
\begin{abstract}
Recently, Memory-based Neural Recommenders (MNR) have demonstrated superior predictive accuracy in the task of sequential recommendations, particularly for modeling long-term item dependencies. However,  typical MNR requires complex memory access operations, i.e., both writing and reading via a controller (e.g., RNN) at every time step. Those frequent operations will dramatically increase the network training time, resulting in the difficulty in being deployed on industrial-scale recommender systems. In this paper, we present a novel general \textbf{Chunk} framework to accelerate MNR significantly. Specifically, our framework divides proximal information units into chunks, and performs memory access at certain time steps, whereby the number of memory operations can be greatly reduced. We investigate two ways to implement effective chunking, i.e., PEriodic Chunk (PEC) and Time-Sensitive Chunk (TSC), to preserve and recover important recurrent signals in the sequence. Since chunk-accelerated MNR models take into account more proximal information units than that from a single timestep, it can alleviate the influence of noise in the user-item interaction sequence to a large extent, and thus improve the stability of MNR. In this way, the proposed chunk mechanism can lead to not only faster training and prediction, but even slightly better results. The experimental results on three real-world datasets (weishi, ml-10M and ml-latest) show that our chunk framework notably reduces the running time (e.g., with up to 7x for training \& 10x for inference on ml-latest) of MNR, and meantime achieves competitive performance.
\end{abstract}

}

\maketitle

\IEEEdisplaynontitleabstractindextext

%
\IEEEpeerreviewmaketitle

\IEEEraisesectionheading{\section{Introduction}\label{sec:introduction}}
\IEEEPARstart{W}{ith} the rapid development of Web 2.0, the speed of data production and streaming has gone up to a great extent. Meanwhile, Internet users can easily access various online products and services, which results in a large amount of action feedback. {The extensive user feedback provides a fundamental information source to build recommender systems}, which assist users in finding  relevant products or items of interest. Since users generally access items in chronological order, the item a user will next interact with may be closely relevant to the accessed items in a previous time window. The literature has shown that it is valuable to consider time information and preference drift for better recommendation performance~\cite{hidasi2015session,quadrana2017personalizing,Tang2018PTS,yuan2019simple,ma2019hierarchical}. In this paper, we focus on the task of sequential  (a.k.a., session-based) recommendation, which is built upon the historical behavior trajectory of users.

A critical challenge for sequential recommendation is to effectively model the preference dynamics of users given the behavior sequence. Among all the existing methodologies, Recurrent Neural Networks (RNN) have become the most prevalent approaches with remarkable success~\cite{hidasi2015session,quadrana2017personalizing}. Different from feedforward networks,  the weights of RNN can be well preserved and updated over time via its internal state, which endows RNN with the ability to process sequence. However, learning vanilla RNN for long-term dependencies remains a fundamental challenge due to the vanishing gradient problem~\cite{bengio1994learning}, and it is noted that long-range user sessions widely exist in real applications. For example, users on TikTok ~\footnote{\url{https://www.tiktok.com/en/}} can watch hundreds of mico-videos in an hour since the average playing time of each video takes only 15 seconds. To model long-term item dependencies for the sequential recommendation problem, previous attempts have introduced  Long Short-Term Memory (LSTM)~\cite{hochreiter1997long} \&  Gated Recurrent Units (GRU) ~\cite{hidasi2015session}, temporal convolutional neural network architecture with dilated layers~\cite{yuan2019simple,fajie2019modeling}, attention machine ~\cite{vaswani2017attention,kang2018self}, and external memory  ~\cite{chen2018sequential,wang2018neural}.

Among these advanced methods, the External Memory Network (EMN)~\cite{sukhbaatar2015end,graves2016hybrid}  mostly resembles human cognitive architecture due to its enhanced external memory mechanism. EMN is composed of a neural controller, e.g., RNN, and the external memory, which can be regarded as an extension of standard RNN, including LSTM \& GRU. Unlike RNN, EMN stores useful past information by external memory rather than a squeezed vector. EMN has shown high potentials in areas, such as visual reasoning~\cite{johnson2017inferring}, question answering~\cite{seo2016query}, natural language processing~\cite{cai2017making}. Since 2018, researchers started to apply it in the field of recommendation to improve the accuracy of existing recurrent models ~\cite{chen2018sequential,wang2018neural,ebesu2018collaborative,huang2018improving}, in the following referred to as  Memory-based Neural Recommenders (MNR).

\begin{figure}[tbp]
	\centering
	\includegraphics[width=8.5cm]{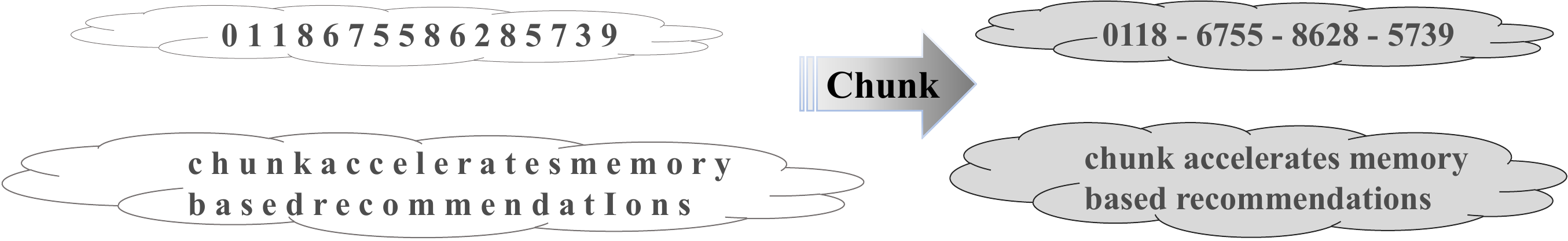}
	\caption{Illustration of the Chunk mechanism for better memorization. Numerics and alphabets are chunked into numeric units and words for faster and easier remembering.}
	\label{chunk}
\end{figure}

In order to remember more information, MNR implementations require to repeat the memory accessing operations, including both reading and writing, at every time step~\cite{sukhbaatar2015end,graves2016hybrid,chen2018sequential,wang2018neural,ebesu2018collaborative,huang2018improving}. The reading and writing accessing operations are much more expensive than the controller in terms of time complexity, which becomes a severe efficiency problem when modeling long-range sequences. One possible way to speed up MNR is to optimize the specific memory operations directly. However, there are many different implementations of accessing operations. \qsl{For example, RUM~\cite{chen2018sequential} updates the entire memory slot iteratively as a first-in-first-out queue, and  NMRN~\cite{wang2018neural} addresses specific memory slots with attention and updates them. It is a challenge to find commonalities in these accessing operations to improve on.} 
In this paper, we focus on developing a general acceleration framework that applies to various types of MNR.  

Our central idea to accelerate MNR in this paper  takes inspiration from the chunk~\cite{Richard2013} technique in cognitive psychology, where  the concept of it is introduced to improve human's memory. Chunk here refers to a meaningful unit of information that can be reorganized based on certain rules. For example, giving the letter sequence "\textbf{\textsl{m-e-m-o-r-y}}", we can remember it as six separate letters, or memorize it by the word "\textbf{\textsl{memory}}", as illustrated in Figure~\ref{chunk}. The latter method can greatly reduce our memory burden but maintain the same amount of information. As such, we believe that applying the chunk strategy for MNR is a promising way to improve the efficiency issue of MNR.

In this paper, we propose a sequential \underline{rec}ommendation framework with \underline{c}hunk-accelerated \underline{m}emory  \underline{n}etwork (CmnRec for short), which speeds up the memory network by reducing the number of memory operations. Our chunk framework consists of the chunk region, chunk rule and attention machine. Specifically, the chunk region temporarily stores the information units (the output vector of the controller) generated in the non-chunk time. The  chunk rule determines when (i.e., chunk time) to perform memory operations. The attention machine extracts the most valuable information in the chunk region, generating new information units to perform memory operations. Through the functions of these modules and rules, chunk compresses the information ingested in the past with high quality, which substantially reduces the workload of memorization so as to improves the recommendation efficiency.

To sum up, the main contributions of this paper include:
\begin{itemize}
	\item {} We propose a general chunk-based sequential recommendation framework, which significantly accelerates various MNRs without harming the accuracy. To the best of our knowledge, this is the first work to evidence that using less memorization can enable comparable accuracy for the recommendation task.
	\item {} \qsl{We present two effective implementations for CmnRec: \textbf{pe}riodic \textbf{c}hunk (PEC) considering the input of each time to be equally important and \textbf{t}ime-\textbf{s}ensitive \textbf{c}hunk (TSC) taking into account both long and short-term dependencies.}
	\item {} We compare CmnRec with state-of-the-art sequential recommendation methods on three real-world datasets. Our experimental\footnote{Our implementation code can be found at \url{https://github.com/SLQu/CmnRec}.} results demonstrate that CmnRec offers competitive and robust recommendations with much less training and inference time.
\end{itemize}

\section{Related Work} 
This work can be regarded as an integration of sequential recommendation and memory networks. In the following, we briefly review related literature in the two directions. 

\subsection{Sequential Recommendation}
Sequential (a.k.a., session-based) recommender systems are an emerging topic in the field of recommendation and have attracted much attention in recent years due to the advance of deep learning. Existing sequential recommendation models can be mainly categorized into three classes according to the  models they involved~\cite{wang2019survey}: Markov chain-based methods~\cite{shani2005mdp,he2016fusing}, factorization based methods~\cite{rendle2010factorizing,yuan2018fbgd,yuan2016lambdafm}, and deep learning-based methods ~\cite{hidasi2015session,hidasi2017recurrent,Tang2018PTS,yuan2019simple}. Specifically, due to the efficiency consideration, Markov chain based recommenders are typically built on the first-order dependency assumption, and thus only capture the first-order dependency over items. As a result, these methods usually do not perform well when modeling long-term and higher-order item dependencies. Factorization-based recommenders (a.k.a., Factorization Machines ~\cite{rendle2010factorizing}) deal with previous user actions as general features by merely summing all their embedding vectors, and are not able to explicitly model the sequential dynamic and patterns in the user session. Thanks to the development of deep neural networks, various deep learning-based sequential models have been proposed and shown superior performance in contrast to the above-mentioned conventional methods by utilizing the complex network architectures.

To be specific, a pioneering work by Hidasi et al.~\cite{hidasi2015session} introduced RNN into the field of recommender systems. They trained a Gated Recurrent Unit (GRU) architecture to model the evolution of user interests, referred to as GRU4Rec. Following this idea, several other RNN variants have been proposed in the past three years. ~\cite{tan2016improved} proposed an improved GRU4Rec by introducing data augmentation and embedding dropout techniques. Hidasi and Karatzoglou~\cite{hidasi2017recurrent} further proposed a family of alternative ranking objective functions with effective sampling tricks to improve the cross-entropy and pairwise ranking losses. ~\cite{quadrana2017personalizing} proposed a personalized sequential recommendation model with hierarchical recurrent neural networks, while ~\cite{gu2016learning, Elena} explored how to leverage content and context features to improve the recommendation accuracy further. Recently, researchers have proposed several other neural network architectures, including convolutional neural networks (CNN) Caser~\cite{Tang2018PTS} and NextItNet~\cite{yuan2019simple}, self-attention models SASRec~\cite{kang2018self}. Compared with RNN models, CNN and attention architectures are much easier to be parallelized on GPUs.

\subsection{EMN and MNR}	
More recently, External Memory Network (EMN) has attracted significant attention in research fields that process sequential data. Generally, EMN involves two main parts: an external memory matrix to maintain state, and a recurrent controller to operate (i.e., reading and writing) the matrix~\cite{chen2018sequential}. Compared with standard RNN models compressing historical signals into a fixed-length vector, EMN is more powerful in dealing with complex relations and long distances due to the external memory. EMN has successfully applied in domains, such as neural language translation~\cite{grefenstette2015learning}, question answering~\cite{miller2016key} and knowledge tracking~\cite{zhang2017dynamic}. Recently, researchers in~\cite{chen2018sequential, ebesu2018collaborative, wang2018neural} have applied it in recommender systems to capture user sequential behaviors and evolving preferences.

As the first work that introduces EMN into the recommendation system, sequential recommendation with user memory network (a.k.a., RUM)~\cite{chen2018sequential} has successfully demonstrated superior advantages over traditional baselines. Similarly, neural memory streaming recommender networks with adversarial training ~\cite{wang2018neural} proposes a  key-value memory network for each user to capture and store both short-term and long-term interests in a unified way. Meanwhile, Ebesu
et al. proposed  collaborative memory network~\cite{ebesu2018collaborative} that deals with all user embedding collections as user memory matrix and utilizes the associative addressing scheme of the memory operations as a nearest neighborhood model.

Ignoring the implementation of external memory networks and the memory operation, all EMN-style models need to perform memory reading and writing operations at every timestep. Such persistent memory operations significantly increase the model complexity and training/inference time, which limits the applications of MNR in large-scale industrial recommender systems. In general, efficiency can be achieved by either reducing the complexity or the frequency of memory access~\cite{le2019learning,Lian2020}. Since there are many ways to implement EMN, we hope to propose a general acceleration framework. To achieve this goal, we propose reducing the number of memory operations, which aims to accelerate MNR with various implementations of memory operations.

\section{Memory-based Neural Recommendation (MNR)}
\label{sec:mnr}
In this section, we will introduce the generic architecture of memory-based sequential recommendation. Let $\mathbb{I}$, $\mathbb{S}$ and $\{x_1,x_2,x_3,...,$ $x_T\}$ (interchangeably denote by $x_{1:T}$) be the set of all items, sequences and items in a specific sequence, respectively. Denote $I=|\mathbb{I}|$ and $S=|\mathbb{S}|$ as the size of item and sequence sets. The corresponding item embedding vectors are $\{\boldsymbol{v}_1,\boldsymbol{v}_2,\boldsymbol{v}_3,...,\boldsymbol{v}_T\}$. 

\begin{figure}[tbp] 
	\center
	\includegraphics[width=8cm]{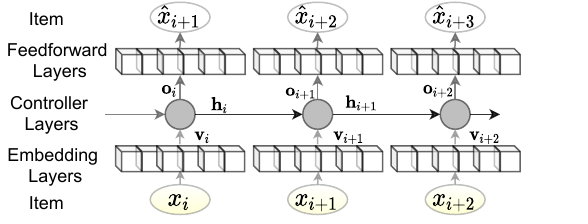}
	\caption{RNN-based Recommendation.}
	\label{CDNC_rnn}
\end{figure}

\qsl{Figure~\ref{CDNC_rnn} is a classical RNN-based recommendation architecture (e.g. GRU4Rec). From bottom to top, the model includes an embedding layer, controller layer, feedforward and the softmax layer of predicted items.}
	
\begin{figure}[tbp] 
	\center
	\includegraphics[width=\linewidth]{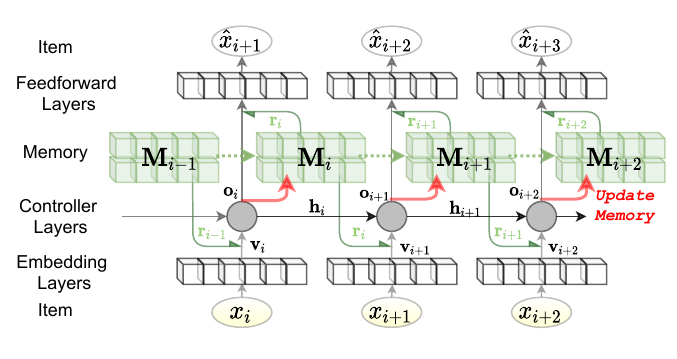}
	\caption{Memory-based Neural Recommendation (MNR).}
	\label{CDNC_m}
\end{figure}

\qsl{Figure~\ref{CDNC_m}  is a generic memory-based neural recommendation architecture. Similar to GRU4Rec, it also  includes embedding, controller, feedforward and the softmax layer. The embedding and controller layers perform exactly the same manner as a classic RNN. The essential difference between MNR and GRU4Rec lies in the memory network layer. In fact, MNR can  also be seen as an extension of RNNs with external memory network $\boldsymbol{M} \in \mathbb{R}^{m*n}$, where $m$ is the number of memory slots and $n$ is the embedding size of memory slot.
}

As shown in Figure~\ref{CDNC_m}, each controller will concatenate the embeddings of the current input item $\boldsymbol{v}_i$ and the memory $\boldsymbol{r}_{i-1}\in \mathbb{R}^n$ (read from $\boldsymbol{M}$) at the previous moment as an external input. The memory storage will be updated according to the output of the controller  $\boldsymbol{o} \in \mathbb{R}^h$. Finally, both the controller output and updated memory $\boldsymbol{r}_{i}$ will be fed into the feedforward layer, which generates the  probabilities of the next interacted item, formulated as follows: 

\begin{equation}
\boldsymbol{f}_i=softmax(\Gamma _{f}(\boldsymbol{o}_i,\boldsymbol{r}_i)) \label{f_layer}
\end{equation}
\begin{equation}
\hat{x}_i = maxID(\boldsymbol{f}_i) \label{max_id}
\end{equation}
where $\Gamma _{f}(\cdot, \cdot)$ is a feedforward operation that performs a linear transformation of the final hidden layer and returns a feature vector $\boldsymbol{f}_i$ as the output. $maxID(f_i)$ is a function to find the item ID with the maximum value in the vector, that is, the maximum occurrence probability at the $i$-th moment predicted by MNR.

\begin{figure*}[tbp] 
	\center
	\includegraphics[width=14cm]{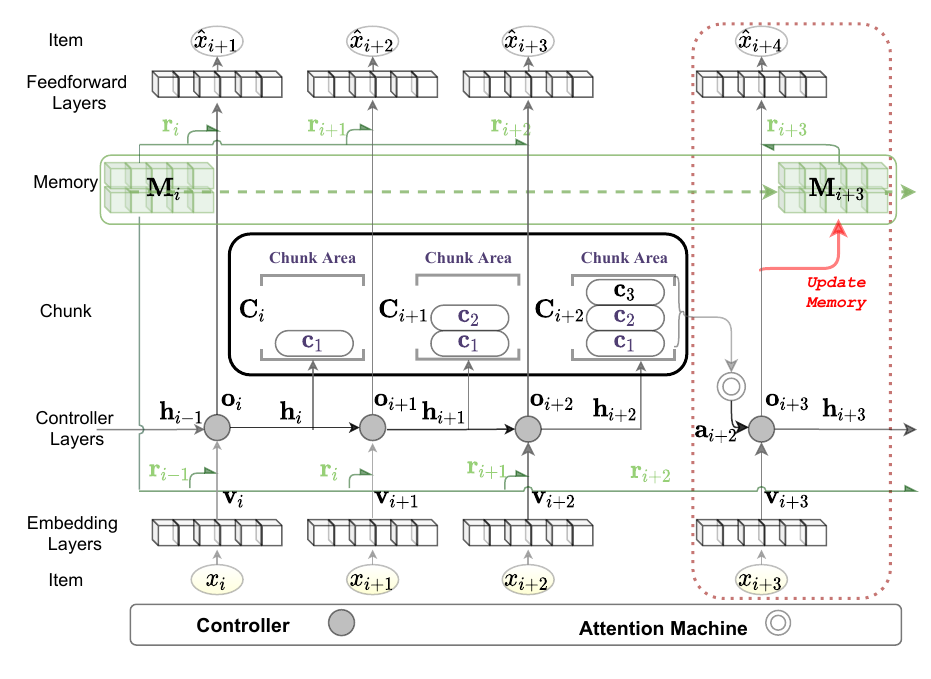}
	\caption{Chunk acceleration on Memory-based Neural Recommendation (MNR). During non-chunk time, information units (the hidden state from the controller) are put into the chunk area. When the chunk time comes, the attention machine will extract the most valuable information in the chunk area and generate a new information unit to replace the current hidden state. Then, the information units stored in the chunk area will be emptied; the memory reading and writing operations are triggered. The dotted red box on the right denotes that MNR performs a complete process from information input and memory update to the final generation of prediction results.}
	\label{CDNC}
\end{figure*}

Let the hidden state of the last moment be $\boldsymbol{h}_{i-1}\in \mathbb{R}^h$. The memory writing operation $\Gamma _{w}(\cdot, \cdot)$ and reading operation $\Gamma _{r}(\cdot, \cdot)$ can be represented as Eq.(\ref{m_update}) and Eq.(\ref{read}):
\begin{equation}
\boldsymbol{M}_i=\Gamma _{w}(\boldsymbol{o}_i,\boldsymbol{M}_{i-1})  \label{m_update}
\end{equation}
\begin{equation}
\boldsymbol{r}_i = \Gamma _{r}(\boldsymbol{o}_i,\boldsymbol{M}_{i}) \label{read}
\end{equation}
The controller output and hidden state are updated as $\Gamma _{o}(\cdot, \cdot)$ and $\Gamma _{h}(\cdot, \cdot)$ :
\begin{equation}
\boldsymbol{o}_i = \Gamma _{o}(\boldsymbol{r}_{i-1},\boldsymbol{h}_{i-1},\boldsymbol{v}_{i},) \label{m_o}
\end{equation}
\begin{equation}
\boldsymbol{h}_i = \Gamma _{h}(\boldsymbol{r}_{i-1},\boldsymbol{h}_{i-1},\boldsymbol{v}_{i},) \label{m_h}
\end{equation}
Normally, $\Gamma _{o}$ and $\Gamma _{h}$ are implemented as RNNs. As for $\Gamma _{w}$ and $\Gamma _{r}$, they have different implementations depending on the selected (external) memory type. In this paper, we adopt the implementations of DNC~\cite{graves2016hybrid} given its  generality.\footnote{\small We refer interested readers to the original paper for detailed explanation due to limited space.}

\section{CmnRec}
In this section, we will give a detailed description of the chunk framework, followed by the concrete implementations.
\subsection{From Psychology to Recommendation}
Psychology points out that people unconsciously use chunk strategies to reduce the ``things'' to be remembered to improve the efficiency of memorization~\cite{Richard2013}. Taken inspiration from this, our central idea of chunk acceleration for MNR is to combine nearby information units according to specific rules and generate new information units so as to reduce the frequency of memory operations and improve memory efficiency. Therefore, how to find an appropriate rule of chunk is the critical question. 

From a more generic perspective, information units are all converted from discrete item sets. An intuitive method is to chunk the information units based on the position of items. In practice, items are  ordered chronologically in the sequence, so the rule of chunk becomes a sequence segmentation problem. That is, how to segment item sequences to minimize information loss while improving memory efficiency?

\subsection{Framework}
The basic idea of chunk-based memory neural network is formed on a specific sequence partitioning rule (described in section \ref{sec:chunk-implementation}), where it first divides the close items into different chunks by order, and then writes these chunks into memory. Suppose the memory slot number of the MNR is $m$, and the length of the sequence is $T$. The whole sequence $x_{1:T}$ will be divided into $m$ subsequences $x_{1:t_1}$, $x_{t_1:t_2}$,..., $x_{t_{m-1}:t_m}$ ($t_m = T$), where $t_1,t_2,...,t_m$ are  chunk time. The controller hidden states  $\boldsymbol{h}$  corresponding to these $m$ subsequences will be chunked $m$ times.

In our chunk framework, the controller output $\boldsymbol{o} $ does not operate memory at every time step, which is the key difference from the standard MNRs~\cite{chen2018sequential,wang2018neural,ebesu2018collaborative,huang2018improving}. To do this, we create a chunk area  $\boldsymbol{C} \in \mathbb{R}^{l*h}$  to store $\boldsymbol{h}$ temporarily. During non-chunk time, $\boldsymbol{C}$ caches $\boldsymbol{h}$. For every $\boldsymbol{h}$ that $\boldsymbol{C}$ caches, $l$ increases by 1. When the chunk time arrives, the attention machine converts $\boldsymbol{C}$ to a new controller hidden state $\boldsymbol{a} \in \mathbb{R}^h$ and then replaces the hidden state in the current controller. Finally, the chunk area is emptied ($l=0$) and memory is manipulated. The formulas of the chunk process are expressed as follows:
\begin{equation}
\boldsymbol{C} = concat(\boldsymbol{h}_{i-l+1},\boldsymbol{h}_{i-l+2},...,\boldsymbol{h}_{i-1},\boldsymbol{h}_{i})\label{chunk_area}
\end{equation}
\begin{equation}
\boldsymbol{z_{ij}}= \boldsymbol{w}tanh(\boldsymbol{W}\boldsymbol{c}_{j}+\boldsymbol{U}\boldsymbol{r}_{i-1}) \label{att_1}
\end{equation}
\begin{equation}
a_{ij} = softmax(\boldsymbol{z_{ij}})  \label{att_3}
\end{equation}
\begin{equation}
\boldsymbol{a}_i=\sum_{j=1}^{l}a_{ij}\boldsymbol{c}_{j}   \label{att_2}
\end{equation}
where $\boldsymbol{c}_{j}$ is the $j$-th element of $\boldsymbol{C}$,  $a_{ij}$ is the attention score of $\boldsymbol{c}_{j}$ at $i$-th time step, and $r_{i-1}$ is the read vector at time step $i-1$. $\boldsymbol{W}\in \mathbb{R}^{b*h}$, $\boldsymbol{U}\in \mathbb{R}^{b*h}$ and  $\boldsymbol{w}\in \mathbb{R}^{b}$ are parameters, where $b$ is attention dimension. And Figure~\ref{CDNC} shows the architecture. Algorithm~1 summarizes the whole process of chunk-enhanced MNR. The theoretical complexity analysis is attached in Appendix \ref{Complexity_analysis}.

\begin{algorithm}
	\caption{CmnRec}
	\KwIn{a original sequence item IDs $x_{1:T-1}$, \\ \quad\quad\quad \, a chunk matrice $\boldsymbol{C}$, \\ \quad\quad\quad \,  a memory slot number $m$.}
	\KwOut{The predicted sequential item IDs  $\hat{x}_{2:T}$}
	Generates chunk time steps set $Ctime$ using Eq.(\ref{tsc});\\
	\For{$i=1;i < T-1;i++$}
	{
		$\boldsymbol{C}.add(h_{i})$   \;
		\texttt{// The operation of the controller and memory in the chunk moment. }\\
		\If{$(i$ in $Ctime)$}
		{
			Use Eq.(\ref{att_1}) (\ref{att_3}) and (\ref{att_2}) to calculate $\boldsymbol{a}_i$\;
			Perform Eq.(\ref{m_o}):                  $\boldsymbol{o}_i = \Gamma _{o}(\boldsymbol{r}_{i-1},\boldsymbol{a}_{i},\boldsymbol{v}_{i})$ \;
			Perform Eq.(\ref{m_h}): $\boldsymbol{h}_i = \Gamma _{h}(\boldsymbol{r}_{i-1},\boldsymbol{a}_{i},\boldsymbol{v}_{i})$ \;
			Perform Eq.(\ref{m_update}): $\boldsymbol{M}_i=\Gamma _{w}(\boldsymbol{o}_i,\boldsymbol{M}_{i-1})$ \;
			Perform Eq.(\ref{read}): $\boldsymbol{r}_i = \Gamma _{r}(\boldsymbol{o}_i,\boldsymbol{M}_{i}) $ \;
			$\boldsymbol{C}.empty()$
		}
		\texttt{// The operation of the controller in the non-chunk moment.}\\
		\Else{			
			Perform Eq.(\ref{m_o}): $\boldsymbol{o}_i = \Gamma _{o}(\boldsymbol{r}_{i-1},\boldsymbol{h}_{i-1},\boldsymbol{v}_{i})$ \;
			Perform Eq.(\ref{m_h}): $\boldsymbol{h}_i = \Gamma _{h}(\boldsymbol{r}_{i-1},\boldsymbol{h}_{i-1},\boldsymbol{v}_{i})$ \;
			$\boldsymbol{r}_i = \boldsymbol{r}_{i-1}$ \;
		}
		\texttt{// Predict the item ID with the highest probability.}\\
		Perform Eq.(\ref{f_layer}): $\boldsymbol{f}_i=softmax(\Gamma _{f}(\boldsymbol{o}_i,\boldsymbol{r}_i))$ \;
		Perform Eq.(\ref{max_id}): $\hat{x}_i = maxID(\boldsymbol{f}_i)$ 
	}
\end{algorithm}

\subsection{Analysis of implementation theory}
In this subsection, we introduce a  metric to evaluate the memorization ability of RNN and Chunk, which helps guide the establishment of appropriate chunk rules.

Let $\boldsymbol{h}_t = \Psi (\boldsymbol{h}_{t-1},\boldsymbol{v}_t) $ be the  transformation formula for hidden states in RNN.
We define the concept of ``contribution'' to measure the strength of the influence, and use the norm of gradient $\left \| \frac{\partial \boldsymbol{h}_{t} }{\partial  \boldsymbol{h}_{t-1}} \right \|$ and $\left \| \frac{\partial \boldsymbol{h}_{t} }{\partial  \boldsymbol{v}_{t}} \right \|$ to represent the contributions of  $\boldsymbol{h}_{t-1}$ and  $\boldsymbol{v}_t$ to  $\boldsymbol{h}_t$. Since the gradient represents the rate of change of trainable variables,  the larger the gradient norm, the greater the contribution it has. As RNN is a cyclic structure, it is also able to measure the contribution of $\boldsymbol{h}_i$ and $\boldsymbol{v}_i$ (to $\boldsymbol{h}_t$) by $\left \| \frac{\partial \boldsymbol{h}_{t} }{\partial  \boldsymbol{h}_{i}} \right \|$ and $\left \| \frac{\partial \boldsymbol{h}_{t} }{\partial  \boldsymbol{v}_{i}} \right \|$. Tersely, let $p_{i,t}$ and $q_{i,t}$ denote these two terms. For standard RNNs, there must be values $\vartheta _p,\vartheta _q \in\mathbb{R}^+ $ that satisfy $\vartheta _p p_{i,t}\geq p_{i-1,t}$ and $\vartheta _q q_{i,t}\geq q_{i-1,t}$ (see Appendix \ref{find_upper_bound} for proof). From past to future, the contribution of $\boldsymbol{v}$ grows when $\vartheta _q < 1$. Correspondingly, we can simply define the total contributions of a sequence with length $t$ in RNN as follows:
\begin{equation}
q_{1,t}+q_{2,t}+...+q_{t,t}  = \sum_{i=0}^{t-1} \vartheta _q^{i} q_{t,t} \label{rnn_con}
\end{equation}

For the chunk-enhanced MNR, the contributions of each chunk area can be counted as a separate RNN contribution. Let the lengths of $m$ chunk areas be $l_1$,$l_2$,..,$l_m$, and $T=\sum_{r=1}^{m} l_r$. Each chunk operation integrates $l$ outputs of the controller.  Hence, the contributions of $m$ chunk areas can be expressed as follows (see Appendix \ref{find_upper_bound} for proof).
\begin{equation}
\underset{m}{\underbrace{  \sum_{i=0}^{l_1 -1} \vartheta _q^{i} \vartheta _p^{T-t_1} ,\sum_{i=0}^{l_2 -1} \vartheta _q^{i} \vartheta _p^{T-t_2} ,\ ...\ ,\sum_{i=0}^{l_m -1} \vartheta _q^{i}\vartheta _p^{T-t_m}  }}   \label{eq:con_3}
\end{equation}
For brevity, let $g_r = \sum_{i=0}^{l_r -1}  \vartheta _q^{i} \vartheta _p^{T-t_r}$, where $r \in \{ 1,2,..,m \}$.
Here $g_r$ contains two parts: (1) $\sum_{i=0}^{l_r -1} \vartheta _q^{i}$ represents the summation of $\boldsymbol{h}_{l_r}$-based contributions in a subsequence, which is considered as the instantaneous contribution. (2) $\vartheta _p^{T-t_r} $ is the proportion between the hidden states $\boldsymbol{h}_{t_r}$ and $\boldsymbol{h}_{t_m}$, which is long-term dependence. These two parts determine the amount of information stored in the memory slot. In order to maximize the total amount of information stored in all memory slots, we should reduce the gaps among $g_1$,$g_2$,... , $g_m $.                                                                                                                                                                                                                                                                                                                                                                                                                                                                                                                                                                                                                                                                                                                                                                                                                          In the following, we introduce two segmentation strategies to achieve this goal. 
\subsection{Chunk Implementation}
\label{sec:chunk-implementation}

	\begin{figure*}[htb]
	\centering
	\subfigure[weishi]{
		\includegraphics[width=5.72cm,height=3.1cm]{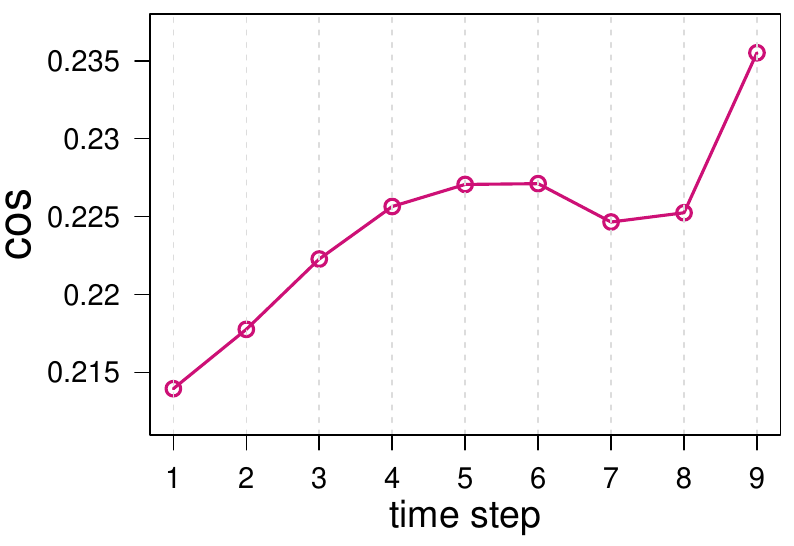}
	}
	\subfigure[ml-10M]{
		\includegraphics[width=5.72cm,height=3.1cm]{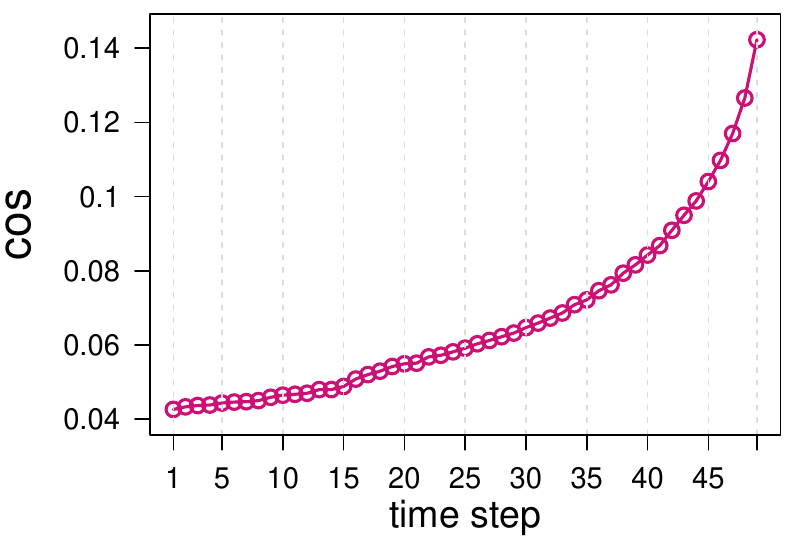}
	}
	\subfigure[ml-latest]{
		\includegraphics[width=5.72cm,height=3.1cm]{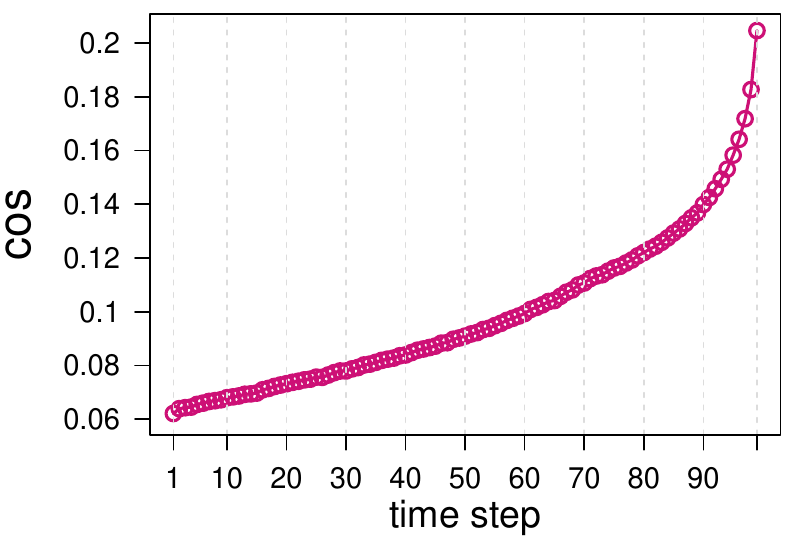}
	}
	\caption{Correlation between the target item and other items in the sequence. (a),(b),(c) represent the correlations on three different datasets, the sequence lengths of which are 10, 50, and 100, respectively.}
	\label{cos}
\end{figure*}

\begin{figure}[tb]
	\centering
	\subfigure[Periodic Chunk (PEC).]{
		\includegraphics[width=5.5cm]{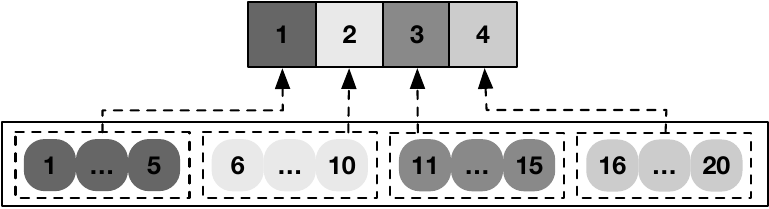}
	}
	\subfigure[Time-sensitive Chunk (TSC)]{
		\includegraphics[width=5.5cm]{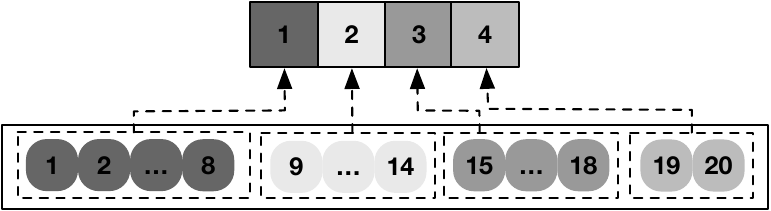}
	}
	\subfigure[Extreme Chunk (EXC)]{
		\includegraphics[width=5.5cm]{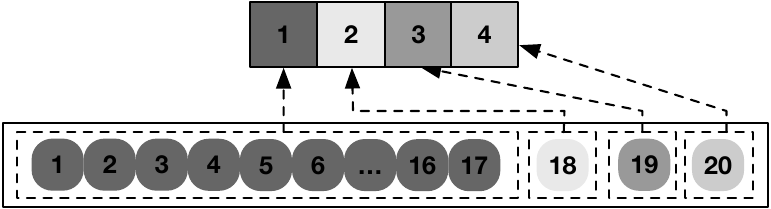}
	}
	\caption{Chunk rule analysis. Given an input item sequence length of 20, the number of memory slots is 4. (a) shows the periodic chunk with period length of 5 and sequence segmentation $x_{1:5},x_{6:10},x_{11:15},x_{16:20}$, updating memory at time steps 5, 10, 15 and 20. (b) shows time-sensitive chunk, where sequence segmentation and memory update time steps are $x_{1:8},x_{9:14},x_{15:18},x_{19:20}$ and 8, 14, 18, 20 respectively. (c) shows the extreme chunk, where the sequence is divided into $x_{1:17},x_{18},x_{19},x_{20}$, and memory is updated at time steps 17, 18, 19 and 20.}
	\label{cra}
\end{figure}

\subsubsection{Periodic Chunk (PEC)}
When the change rate of $\boldsymbol{v}$ in the sequence is relatively slow, i.e., the preference transfer of users is not remarkable, we have $\vartheta _p \rightarrow 1$ and $\vartheta _q \rightarrow 1$. This means the long-term dependencies of each chunk are similar, and the instantaneous contribution increases with the length of the subsequence. Only when all instantaneous contributions are the same (i.e., $l_1=l_2=...=l_m$), the gaps among each chunk contribution are the smallest. Therefore, we propose the periodic chunk (PEC). Given the input sequence $x_{1:T}$ and the chunk cycle $G= \left \lfloor \frac{T}{m} \right \rfloor $,  the chunk time steps are:
\begin{equation}
\underset{m \; chunk \; time \; steps.}{\underbrace{T- \ (m-1) \ G,T \ -\  (m-2) \ G,\ ...\ ,\ T-2G,\ T-G,\ T}}
\end{equation}
The sequence segmentation results are:
\begin{equation}
\begin{aligned}
&x_{1:T\ -\ (m-1)\ G},\ x_{T\ -\ (m-1)\ G:T\ -\ (m-2)G}, \\
&\ ...\ , \ x_{T\ -\ 2G:T\ -\ G}, \ x_{T\ -\ G:T}
\end{aligned}
\end{equation}
Figure \ref{cra} (a) is a graphical illustration of PEC. 

\textbf{Analysis.} In the long run, user preferences always shift, and users usually have different degrees of preference to different items, which means there may be a distribution of user preferences in the sequence. To demonstrate the preference distribution, we investigate the importance of items to the target item in a sequence. In a given sequence, the last item is treated as target item. We calculated the item importance as the correlation between the current item in the sequence and the target item. Specifically, we  adopt cosine similarity as the correlation indicator, which is given as follows.
\begin{equation}
cosine( \boldsymbol{v}_i, \boldsymbol{v}_j)=\frac{\boldsymbol{v}_i\cdot \boldsymbol{v}_j}{\left \| \boldsymbol{v}_i \right \|\left \|  \boldsymbol{v}_j \right \|} =\frac{\sum_{r=1}^{k}v_{ir}v_{jr}}{\sqrt{\sum_{r=1}^{k}v^2_{ir}} \sqrt{\sum_{r=1}^{k}v^2_{jr}}}      
\end{equation} 
We use the item embeddings (trained by the controller) as the input vectors for cosine similarity.
Experimental results are shown in Figure~\ref{cos}, the horizontal axis is the position of items in a sequence, and the vertical axis is the correlation between the target item and the current item. As shown, although there are some small fluctuations in Figure \ref{cos} (a), the overall trend of the correlation in  (b) (c) and (d) is stable and upward. These figures generally indicate that the correlations between a target user interaction and previous interactions increase as the their interval time  decrease. That is, the newer the interaction is, the more it reflects future changes of preference in the sequence. 

\subsubsection{Time-sensitive Chunk (TSC)}
\label{sec:Tsc}
Inspired by the above analysis, it is reasonable to propose a time-sensitive chunk strategy where the writing interval is larger at the beginning of the interaction sequence but will be reduced over time  (i.e., $l_1<l_2<...<l_m$), so as to enhance the impact of latest user interactions. As the input length between each chunk is $1, 2, .., m-1$ and $m$,  the sum of the input length is $\frac{m(m+1)}{2}$, and proportional step is $g= \left \lfloor \frac{2T}{m(m+1)} \right \rfloor $. Hence, the chunk time steps are given as:
\begin{equation}
\underset{m \; chunk \; time \; steps}{\underbrace{T-g\frac{2T}{m(m-1)},T-g\frac{2T}{(m-1)(m-2)},...,T-g,T}} \label{tsc}
\end{equation}
The sequence segmentation results are:
\begin{equation}
\begin{aligned}
x_{1:T\ -\ g\ \frac{2T}{m\ (m-1)}},x_{T\ -\ g\ \frac{2T}{m\ (m-1)}:T\ -\ g\ \frac{2T}{(m-1)(m-2)}},...,x_{T\ -\ g,T}
\end{aligned}
\end{equation}
The example of TSC is shown in Figure \ref{cra} (b). To obtain the goal of  ``the newer the interaction is, the greater importance it has for the next prediction", we may need to investigate an extreme case, as shown in Figure \ref{cra} (c), referred to as EXC. As can be seen, in EXC the most attention is paid to the contribution of latest interaction, i.e., the first $T-m+1$ items form a large chunk, whereas each of the remaining $m-1$ items is treated as a separate chunk. 

\subsection{Model Discussion}
We conduct model analysis to identify the difference between CmnRec  and two related models RUM~\cite{chen2018sequential} and NextItNet~\cite{yuan2019simple}. 
\subsubsection{Relation with RUM} 	
RUM is considered as the first work that adapts external memory network (EMN) for the recommendation task. By introducing an EMN, its ability to model long sequences has been greatly improved. Figure~\ref{architectures}(a) illustrates the structure of EMN. As can be seen, both RUM and CmnRec rely on the reading and writing operations when using and updating memory. However, RUM needs to perform these operations at every time step. Such frequent and continuous memory operations largely increase the computation costs and cannot robustly process noisy information. By contrast, CmnRec only needs to operate at specific time steps. By caching the input information during non-memory update time and using the attention machine to extract the valuable information during memory update, CmnRec achieves efficient and high-quality recommendation results. 

\subsubsection{Relation with NextItNet}	
\begin{figure}[tb]
	\centering
	\subfigure[RUM]{
		\includegraphics[width=4.0cm]{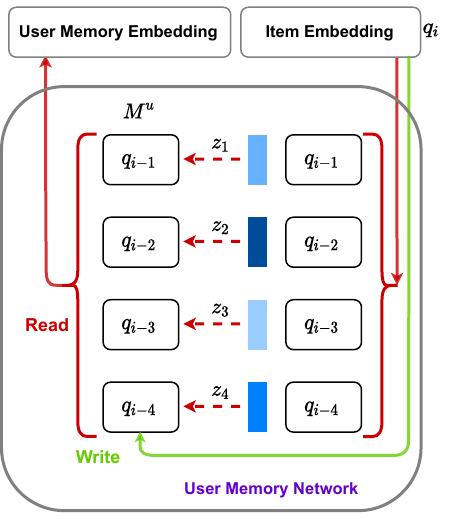}
	}
	\subfigure[NextItNet]{
		\includegraphics[width=4.0cm]{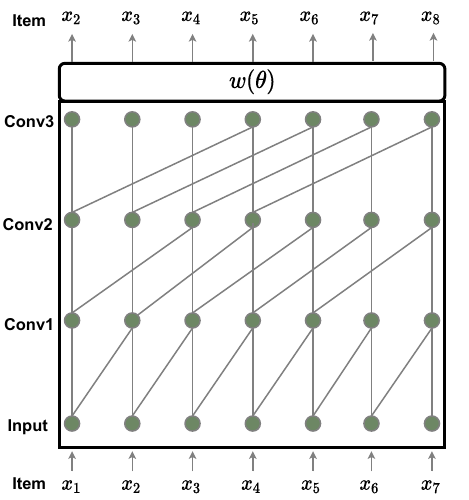}
	}
	\caption{User memory network of RUM and architecture of NextItNet.}
	\label{architectures}
\end{figure}

NextItNet is a classic application of dilated CNNs for the sequential recommendation task. By puncturing regular  holes on the convolutional kernel, NextItNet is endowed higher capacity to remember longer sequence without increasing the kernel size or stacking more convolutional layers. The network architecture of NextItNet is illustrated in Figure~\ref{architectures}(b) with kernel size of  $1*2$. From the memory saving perspective, the idea of CmnRec and NextItNet is similar in some ways since both models perform connection skipping to reduce memory usage. The main difference lies in that NextItNet performs convolutions and memorization with fixed time intervals, while CmnRec performs memorization according to the chunk rules and has no such a restriction as mentioned before. In addition, both the controller and external memory of CmnRec have the memorization capacity, while NextItNet does not have a specific memory container for memorization. 

	\section{Experiments}
In this section, we  conduct extensive experiments to investigate the efficacy of the
chunk-accelerated MNR. Specifically, we aim to answer the following research questions (RQs).

\begin{enumerate}
	\item \textbf{RQ1:} 
	Does chunk speed up MNR significantly? What impacts does the sequence length have on model acceleration?
	\item \textbf{RQ2:}  Does  the chunk-accelerated MNR perform comparably with  the typical memory-based neural recommendation models in terms of recommendation accuracy?
	\item \textbf{RQ3:}  How does chunk-accelerated MNR perform with  TSC, EXC and PEC? Which setting performs best? 
\end{enumerate}

\subsection{Datasets}
We conduct experiments on three real-world recommendation datasets:  ml-latest, ml-10M\footnote{https://grouplens.org/datasets/movielens/} and weishi\footnote{https://www.weishi.com/}.             

\noindent\textbf{ml-latest}~\cite{harper2016movielens} is a widely used public dataset for both general and sequential recommendations~\cite{tang2019towards,fajie2019modeling,wang2019towards,kang2018self}. The original dataset contains 27,753,444 interactions, 283,228 users and 58,098 video clips with timestamps. To reduce the impact of cold items, we filter out videos that appear less than 20 times, and  generate a number of  sequences, each of which belongs to one user in chronological order. Then, we split  the sequence into subsequence every $L$ movies. If the length of the subsequence is less than $L$, we pad zero in the beginning of the sequence to reach $L$. 
For those with length less than $l$, we simply remove them in our experiments. In our experiments, we set $L=100$ with $l=20$.

\begin{table}[tp]
	\scriptsize
	\centering
	\caption{The statistics of the experimental datasets. \textsl{s}: the average length of each sequence. \textsl{T}: the unified sequence length after padding zero.}
	\label{datasets}
	\begin{tabular}{cccccc}
		\toprule
		Dataset & \# Interactions& \# Sequence& \# Item &\textsl{s}&\textsl{T}\\
		\midrule
		weishi&9,986,953&1,048,575&65,997&9.5243&10\\
		ml-10M&7,256,224&178,768&10,670&40.5902&50\\
		ml-latest&25,240,741&300,624&18,226&83.9612&100\\
		\bottomrule
	\end{tabular}
\end{table}

\begin{table}[tp]
	\caption{Inference speedup. The values denote multiples.  m is slot number.}
	\scriptsize
	\centering
	\label{tab:time_reduce}
	\begin{tabular}{c|cccccc|c}
		\toprule
		m&2&3&4&6&9&12&Average  \\
		\midrule
		weishi&1.51&1.52&--&--&--&--&1.515  \\
		ml-10M&6.28&--&6.71&6.17&4.68&--&5.96  \\
		ml-latest&11.35&--&12.16&10.51&8.28&8.03&10.07\\
		\bottomrule
	\end{tabular}
\end{table}
\noindent	\textbf{ml-10M} 
contains 10,000,054 interactions, 10,681 movies and 71,567 users. We perform similar pre-processing as ml-latest by setting $L$ to 50 and $l$ to 5.

\noindent	\textbf{weishi} is a micro-video recommendation dataset collected by the Weishi Group of Tencent. Since both cold users and items have already been trimmed by the official provider, we do not need to perform pre-processing for the cold-start problem. Each user sequence contains 10 items at maximum. The statistics of our datasets after above preprocessing are shown in Table ~\ref{datasets}.

\subsection{Comparative Methods \& Evaluation Metrics }
\label{sec:comparison-methods}
\textbf{GRU4Rec}~\cite{hidasi2015session}: It is  a seminal work that applies the Gated Recurrent Unit (GRU) for sequential recommendation. 
For a fair comparison, we use the cross-entropy loss function for all neural network models. \textbf{LSTM4Rec}: It simply
replaces GRU with LSTM since we observe that LSTM generally performs better than GRU for the item recommendation task.  \textbf{SRMN}~\cite{chen2018sequential}: It is a recently proposed sequential recommendation model with external memory network architecture. For comparison purpose, we report results by using LSTM as the controller. In addition, we also compare with two CNN-based sequential recommendation methods:~\textbf{Caser}~\cite{Tang2018PTS} and \textbf{NextItNet}~\cite{yuan2019simple}. 
As for our proposed methods, we report results with the three chunk variants, i.e., TSC, PEC and EXC. Note that following the original paper of GRU4Rec and NextItNet, we conduct all our experiments without learning an explicit user embedding.\footnote{We empirically found that concating a
	user embedding vector (e.g., in Caser) for sequential recommendation models do not yield any better results. This is probably because the user embedding has already been well represented by the
	embedding of his interaction sequences, as well illustrated in \cite{kang2018self}. 
}  

Following~\cite{yuan2019simple,Tang2018PTS}, we use three popular top-$N$ metrics to evaluate the performance of these sequential recommendation models, namely, MRR@\textit{N} (Mean Reciprocal Rank)~\cite{hidasi2017recurrent}, HR@\textit{N} (Hit Ratio)~\cite{wang2018neural} and NDCG@\textit{N} (Normalized Discounted Cumulative Gain)~\cite{guo2016personalized}. 
\begin{figure*}[htbp]
	\centering
	\subfigure[weishi]{
		\includegraphics[width=5.72cm,height=3.9cm]{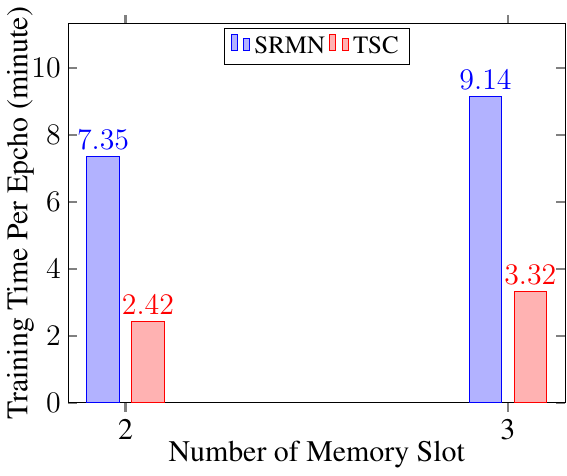}
	}
	\subfigure[ml-10M]{
		\includegraphics[width=5.72cm,height=3.9cm]{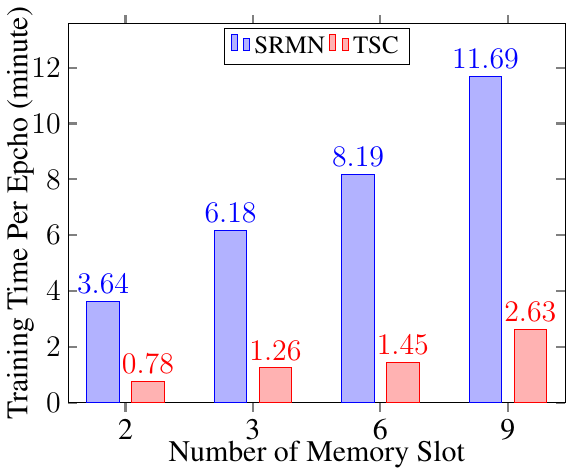}
	}
	\subfigure[ml-latest]{
		\includegraphics[width=5.72cm,height=3.9cm]{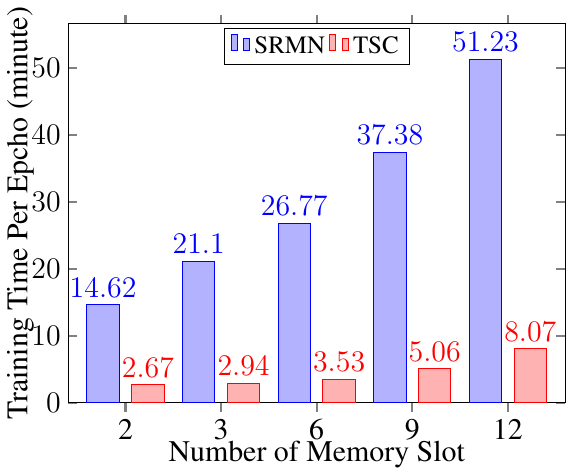}
	}
	\caption{Training time of each epoch on the three datasets. }
	\label{timereduce}
\end{figure*}

\subsection{Experiment Setup }
To ensure the fairness of the experiment, the  dimensions of item embeddings are set to 128 for all neural network models, similar to~\cite{Tang2018PTS,yuan2019simple}. We first tune baseline GRU4Rec and LSTM4Rec to optimal performance. Specifically, we set the number of layers of GRU4Rec and LSTM4Rec to 1 and the hidden dimension to 256, which performs better than two hidden layers or a  larger hidden dimension. We empirically find that all models except NextItNet benefit from a larger batch size. To make full use of GPU, we set batch size to 1024 for these models. As for NextItNet, we empirically find that it performs best when batch size is between 64 and 256 for all these datasets.	We report the results with its best-performing batch size.  For SRMN, we set the embedding size of memory slot as 256. The attention dimension $b$ of chunk is 64 on all datasets. Our datasets are randomly divided into training (80\%), validation (2\%) and testing (18\%) sets. All methods are implemented using Tensorflow with Adam~\cite{kingma2014adam} as the optimizer. Results are reported when models are converged on the validation test. 
\begin{figure*}[t]
	\centering
	\subfigure[MRR@5]{
		\includegraphics[width=5.72cm]{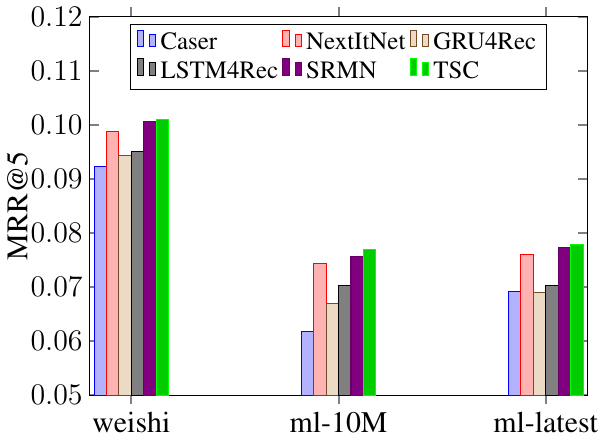}
	}	
	\subfigure[HR@5]{
		\includegraphics[width=5.72cm]{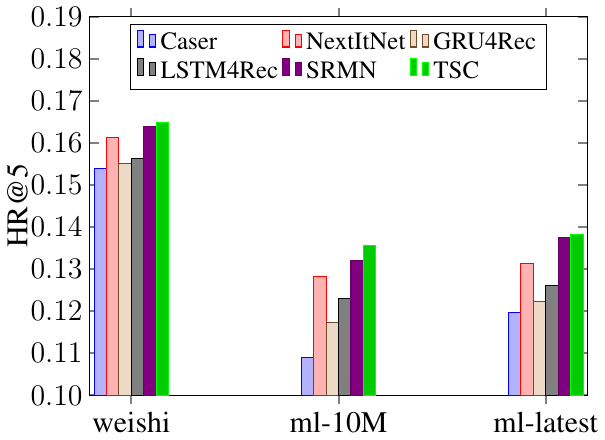}
	}
	\subfigure[NDCG@5]{
		\includegraphics[width=5.72cm]{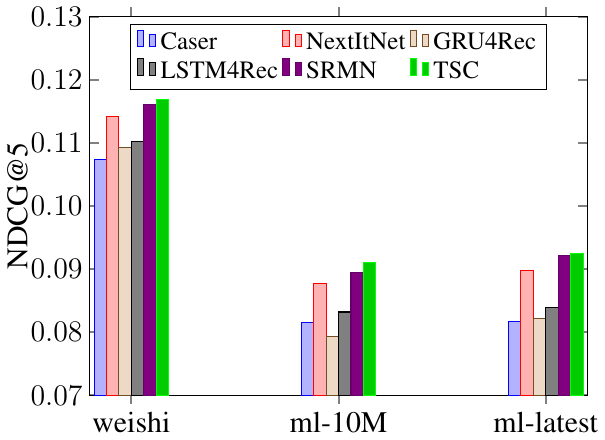}
	}
	\caption{Performance comparisons with respect to top-N values. }
	\label{baselines}
\end{figure*}

\begin{table*}
	\caption{Performance comparison between SRMN and the proposed methods. Bold means the best result, $*$ means the second-best result. $m$  is slot number. 
	}
	\centering
	\label{tab:evaluate1}
	\scriptsize
	\begin{tabular}{c|c|cc|cccc|ccccc}
		\toprule
		\multicolumn{2}{c|}{Dataset} &\multicolumn{2}{c|}{weishi}&\multicolumn{4}{c}{ml-10M}&\multicolumn{5}{c}{ml-latest}\\ 
		\midrule
		\multicolumn{2}{c|}{$m$}&2&3&2&4&6&9&2&4&6&9&12\\            
		\midrule\midrule
		\multirow{4}{*}{\rotatebox{90}{MRR@5}}
		&SRMN&\textbf{0.1001} &0.1005 &\textbf{0.0739} &\textbf{0.0748} &$0.0751^*$ &0.0756&\textbf{0.0741} &\textbf{0.0749} &\textbf{0.0755} &\textbf{0.0764} &$0.0773^*$ \\
		&TSC&0.0978 &\textbf{0.1010} &$0.0727^*$ &0.0724 &0.0738 &\textbf{0.0769}&$0.0721^*$ &0.0733 &0.0742 &$0.0755^*$  &\textbf{0.0778} \\
		&PEC&$0.0985^*$ &$0.1006^*$ &0.0721 &$0.0731^*$ &\textbf{0.0753} &$0.0768^*$ &0.0717 &$0.0734^*$ &$0.0753^*$ &0.0750&0.0750 \\
		&EXC&0.0958 &0.0954 &0.0651 &0.0633 &0.0641 &0.0673 &0.0636 &0.0593 &0.0622 &0.0633 &0.0653 \\
		\midrule
	\end{tabular}
	
	\begin{tabular}{c|c|cc|cccc|ccccc}
		\multirow{4}{*}{\rotatebox{90}{HR@5}}
		&SRMN&\textbf{0.1636} &0.1638&\textbf{0.1302} &\textbf{0.1316} &$0.1318^*$&0.1320&\textbf{0.1317} &\textbf{0.1335} &\textbf{0.1340} &\textbf{0.1359} &$0.1374^*$ \\
		&TSC&0.1599 &\textbf{0.1648} &$0.1285^*$ &
		0.1287 &0.1302 &\textbf{0.1356} &0.1289 &0.1318 &0.1329 &$0.1348^*$ &\textbf{0.1381} \\
		&PEC&$0.1607^*$ &$0.1640^*$ &0.1280 &$0.1305^*$ &\textbf{0.1336} &$0.1350^*$  &$0.1300^*$ &$0.1321^*$ &$0.1345^*$ &0.1346&0.1347 \\
		&EXC&0.1573 &0.1567 &0.1157 &0.1125 &0.1137 &0.1188 &0.1149 &0.1088 &0.1132 &0.1161 &0.1190 \\
		\midrule
	\end{tabular}

	\begin{tabular}{c|c|cc|cccc|ccccc}
		\multirow{4}{*}{\rotatebox{90}{NDCG@5}}
		&SRMN&\textbf{0.1158}&0.1161&\textbf{0.0878}&\textbf{0.0888}&$0.0890^*$&0.0895&\textbf{0.0883}&\textbf{0.0893}&\textbf{0.0899}&\textbf{0.0911}&$0.0921^*$ \\			
		&TSC&0.1131&\textbf{0.1168}&$0.0864^*$&0.0863&0.0876&\textbf{0.0911}&0.0860&0.0875&0.0882&$0.0901^*$&\textbf{0.0925} \\ 
		&PEC&$0.1139^*$&$0.1162^*$&0.0857&$0.0871^*$&\textbf{0.0895}&$0.0909^*$&$0.0860^*$&$0.0878^*$&$0.0897^*$&0.0897&0.0896 \\
		&EXC&0.1110&0.1105&0.0774&0.0754&0.0763&0.0800&0.0758&0.0713&0.0746&0.0762&0.0784 \\
		
		\bottomrule
	\end{tabular}
\end{table*}

\begin{table}
	\caption{\qsl{Performance comparisons against baseline with respect to top-5 values. $\uparrow$ means the percentage of TSC's performance over baseline methods}}
	\scriptsize
	\centering
	\label{tab:evaluate_baseline_5}
	\begin{tabular}{c|c|llll}
		\toprule
		\multicolumn{2}{c|}{Dataset} &\multicolumn{1}{c}{weishi}&\multicolumn{1}{c}{ml-10M}&\multicolumn{1}{c}{ml-latest}\\   
		\midrule
		\multirow{7}{*}{\rotatebox{90}{MRR@5}}
		&Caser    &0.0922($\uparrow$ 8.71\%)&0.0618($\uparrow$ 19.63\%)&0.0602($\uparrow$ 22.62\%)\\
		&NextItNet&0.0987($\uparrow$ 2.28\%)&0.0743($\uparrow$ 3.34\%)&0.0761($\uparrow$ 2.19\%)\\	
		&GRU4Rec  &0.0943($\uparrow$ 6.63\%)&0.0670($\uparrow$ 12.87\%)&0.0690($\uparrow$ 11.31\%)\\
		&LSTM4Rec &0.0950($\uparrow$ 5.94\%)&0.0702($\uparrow$ 8.71\%)&0.0703($\uparrow$ 9.64\%)\\
		&SRMN      &0.1005($\uparrow$ 0.50\%)&0.0756($\uparrow$ 1.69\%)&0.0773($\uparrow$ 0.64\%)\\
		&\textbf{TSC}  &\textbf{0.1010}&\textbf{0.0769}&\textbf{0.0778}\\
		\midrule
		\multirow{7}{*}{\rotatebox{90}{Hit@5}}
		&Caser    &0.1538($\uparrow$ 6.67\%)&0.1089($\uparrow$ 19.69\%)&0.1198($\uparrow$ 13.25\%)\\
		&NextItNet&0.1613($\uparrow$ 2.12\%)&0.1283($\uparrow$ 5.38\%)&0.1314($\uparrow$ 4.85\%)\\	
		&GRU4Rec  &0.1550($\uparrow$ 5.59\%)&0.1172($\uparrow$ 13.57\%)&0.1223($\uparrow$ 11.44\%)\\
		&LSTM4Rec &0.1562($\uparrow$ 5.22\%)&0.1230($\uparrow$ 9.29\%)&0.1261($\uparrow$ 8.69\%)\\
		&SRMN      &0.1638($\uparrow$ 0.61\%)&0.1320($\uparrow$ 2.65\%)&0.1374($\uparrow$ 0.51\%)\\
		&\textbf{TSC}  &\textbf{0.1648}&\textbf{0.1356}&\textbf{0.1381}\\
		\midrule
		\multirow{7}{*}{\rotatebox{90}{NDCG@5}}
		&Caser    &0.1074($\uparrow$ 8.05\%)&0.0816($\uparrow$ 10.43\%)&0.0817($\uparrow$ 11.68\%)\\
		&NextItNet&0.1142($\uparrow$ 2.23\%)&0.0877($\uparrow$ 3.73\%)&0.0898($\uparrow$ 2.92\%)\\	
		&GRU4Rec  &0.1092($\uparrow$ 6.51\%)&0.0793($\uparrow$ 12.95\%)&0.0822($\uparrow$ 11.14\%)\\
		&LSTM4Rec &0.1102($\uparrow$ 5.65\%)&0.0832($\uparrow$ 8.67\%)&0.0839($\uparrow$ 9.30\%)\\
		&SRMN      &0.1161($\uparrow$ 0.60\%)&0.0895($\uparrow$ 1.67\%)&0.0921($\uparrow$ 0.43\%)\\
		&\textbf{TSC}  &\textbf{0.1168}&\textbf{0.0911}&\textbf{0.0925}\\
		\bottomrule
	\end{tabular}
	
\end{table}

\begin{table}
	\caption{\qsl{Performance comparisons against baseline with respect to top-20 values.}}
	\scriptsize
	\centering
	\label{tab:evaluate_baseline_20}
	\begin{tabular}{c|c|llll}
		\toprule
		\multicolumn{2}{c|}{Dataset}&\multicolumn{1}{c}{ml-10M}&\multicolumn{1}{c}{ml-latest}\\   
		\midrule
		\multirow{7}{*}{\rotatebox{90}{MRR@20}}
	&Caser&0.0786($\uparrow$ 16.03\%)&0.0755($\uparrow$ 23.44\%)\\
	&NextItNet&0.0879($\uparrow$ 3.75\%)&0.0898($\uparrow$ 3.79\%)\\
	&GRU4Rec&0.0799($\uparrow$ 14.14\%)&0.083($\uparrow$ 12.29\%)\\
	&LSTM4Rec&0.0842($\uparrow$ 8.31\%)&0.0845($\uparrow$ 10.29\%)\\
	&SRMN&0.0905($\uparrow$ 0.77\%)&0.0925($\uparrow$ 0.75\%)\\
	&\textbf{TSC} &\textbf{0.0912}&\textbf{0.0932}\\
	\midrule
	\multirow{7}{*}{\rotatebox{90}{Hit@20}}
	&Caser&0.2373($\uparrow$ 22.37\%)&0.2436($\uparrow$ 22.53\%)\\
	&NextItNet&0.2713($\uparrow$ 7.04\%)&0.275($\uparrow$ 8.54\%)\\
	&GRU4Rec&0.2559($\uparrow$ 13.48\%)&0.2722($\uparrow$ 9.66\%)\\
	&LSTM4Rec&0.272($\uparrow$ 6.76\%)&0.278($\uparrow$ 7.37\%)\\
	&SRMN&0.2829($\uparrow$ 2.65\%)&0.2945($\uparrow$ 1.35\%)\\
	&\textbf{TSC} &\textbf{0.2904}&\textbf{0.2985}\\
	\midrule
	\multirow{7}{*}{\rotatebox{90}{NDCG@20}}
	&Caser&0.1102($\uparrow$ 21.76\%)&0.1127($\uparrow$ 22.11\%)\\
	&NextItNet&0.1279($\uparrow$ 4.92\%)&0.1301($\uparrow$ 5.79\%)\\
	&GRU4Rec&0.1179($\uparrow$ 13.81\%)&0.1238($\uparrow$11.21\%)\\
	&LSTM4Rec&0.1247($\uparrow$ 7.63\%)&0.1261($\uparrow$ 9.13\%)\\
	&SRMN&0.1303($\uparrow$ 2.98\%)&0.1363($\uparrow$ 1.02\%)\\
	&\textbf{TSC} &\textbf{0.1342}&\textbf{0.1377}\\
		\bottomrule
	\end{tabular}
	
\end{table}

\subsection{Experimental Result and Analysis}
\subsubsection{Run time (RQ1).}
As analyzed before, the chunk framework is theoretically more efficient than SRMN by reducing the number of memory access. To confirm this, we plot the results of the running time of the two methods in Figure~\ref{timereduce}. It can be seen that the training time of SRMN is several times slower than that of TSC, and the speedup with the maximum memory slot (best accuracy for both SRMN and TSC) on the three datasets  are 2.75, 4.44, and 6.34 respectively. We find that the relative improvements are much larger on ml-10M and ml-latest than on weishi. The larger improvements should be attributed to the lengths of the item sequence since for longer sequences, the interval distance between two memory accesses is also larger.	Taking the weishi and ml-latest as an example. By setting the number of memory slots as 2, the average interval distance to perform memory access on weishi is 5, while it is 50 on ml-latest. It is also worth noting that the relation between the number of memory slots and the running time is not linear. Increasing the number of memory slots will lead to a decrease of the chunk area, which helps to reduce the computing time of  the attention machine. Therefore, the optimal slot number depends on the specific dataset. We also demonstrate the speedup for item generating in Table \ref{tab:time_reduce}. As shown, similar conclusions also hold to the inference phase. 

\subsubsection{Performance comparison with original SRMN (RQ2)}
To verify the effectiveness of the proposed chunk framework, we focus on comparing it with the standard SRMN. We report the recommendation accuracy on Table \ref{tab:evaluate1}, \qsl{and all models in the same column share the same hyperparameter Settings. The observations are }(1) TSC achieves comparable results with SRMN on all datasets by applying for a relatively large slot number. Both SRMN and our method are senstive to the number of slots --- better accuracy is obtained with larger slot number. Particularly, TSC and PEC with 9 memory slots had an even 1.72\%  performance improvement over SRMN on ml-10M in term of MRR@5. (2) In general, the performance of all chunk-based methods will keep growing  by increasing the number of memory slots in the beginning. It then keeps relatively stable once the number of memory slots has been large enough. The optimal number can be achieved by hyperparameter tuning. Empirically, for a sequential dataset with session length longer than 50, we can set the default number to 10, which is a favorable trade-off between the performance and computational cost.
\subsubsection{Performance comparison against baselines.}
We report the results of all methodologies in Figure ~\ref{baselines} and Table ~\ref{tab:evaluate_baseline_5}, \ref{tab:evaluate_baseline_20}, and make the following observations. First, the CNN-based model Caser performs worse than GRU4Rec and LSTM4Rec. By contrast, the state-of-the-art temporal CNN model NextItNet yields obviously better results than these baselines. Our findings here are consistent with those in previous works~\cite{yuan2019simple,tang2019towards}. Third,  SRMN and TSE outperform all other baselines, which demonstrates the effectiveness of memory-based neural networks. \qsl{Finally, TSC outperforms SRMN in all data sets, which proves the rationality of chunk policy.}
\begin{figure*}[t]
	\centering
	\subfigure[weishi]{
		\includegraphics[width=0.9\columnwidth]{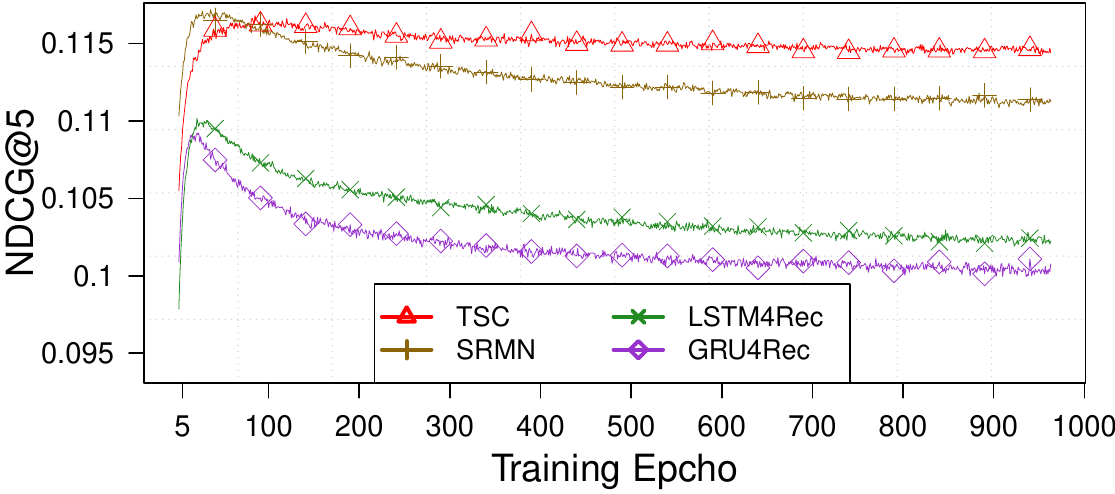}
	} \quad
	\subfigure[ml-10M]{
		\includegraphics[width=0.9\columnwidth]{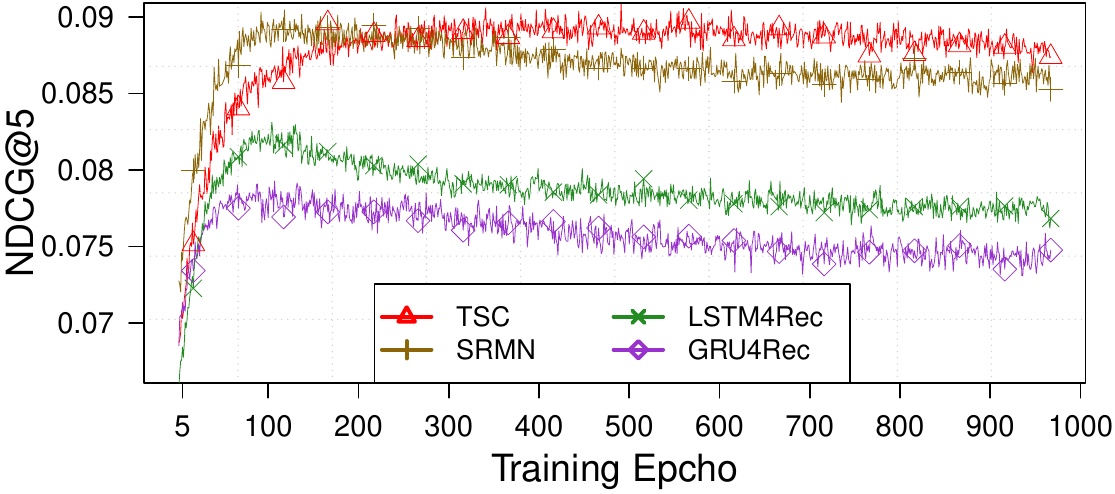}
	}
	\caption{Convergence behaviors in terms of NDCG@5.
		The number of memory slots of TSC and SRMN are 3 and 9 respectively on the two datasets. }
	\label{ndcg5}
\end{figure*}
\subsubsection{Denoising}
We plot the convergence behaviors of GRU4Rec, LSTM4Rec, SRMN and TSC in Figure~\ref{ndcg5}. As shown, memory-based recommendation models (i.e., SRMN and TSC) have apparent advantages over the RNN models in terms of both accuracy and robustness.
We believe the external memory network can enhance the storage and the capacity of information processing of RNN so as to improve accuracy. In addition, abnormal  input data or noise usually leads to overfitting after convergence. However, since the external storage network maintains more information than the recurrent unit, the impacts of abnormal data from a small number of instances
can be restricted to a certain extent. Furthermore, we observe that TSC is even more robust than SRMN, and the results on both weishi and ml-10M imply that TSC can effectively prevent the overfitting problem. We argue that the memory update mechanism in TSC makes it insensitive to noise since it takes into account data obtained from previous timesteps rather than that from only the current timestep. 

\subsubsection{Performance comparison of TSC, PEC and EXC (RQ3)} 
Since we have introduced three chunk variants, i.e., TSC, PEC and EXC, we report their results on Table \ref{tab:evaluate1} for a clear comparison. First, we observe that PEC and TSC perform much better than EXC on all datasets. In fact, EXC performs even worse than the baseline models. We suspect that this is because EXC mainly focuses on modeling  the most recent interactions, ignoring earlier interactions which however make up the vast majority of the interaction sequence. That is, the extreme partitioning cannot offer satisfied performance in practice. Second,  TSC achieves better results than PEC in terms of all evaluation metrics when setting a large slot number. This implies that the time-sensitive chunk strategy is better suited to balance long short-term  sequential relations than the periodic setup. 

\section{Conclusion}
In this paper,  we have introduced a novel sequential recommendation framework by combining the Chunk and External Memory Network (EMN). The motivation is that the way of memory access operations in the existing EMN introduces redundant computation, which results in very high time complexity when modeling long-range user session data. A Chunk-accelerated memory network is proposed with two practical implementations: periodic chunk (PEC) and time-sensitive chunk (TSC). We demonstrate that our proposed chunk framework significantly reduces the computation time of memory-based sequential recommendation models but achieves competitive recommendation results.


%

\appendices
\section{\label{Complexity_analysis}A rough complexity analysis}
MNR consists of a controller and an external memory network (EMN). For easier illustration, the controller and EMN are often realized by RNNs and DNC. And the time complexity of them is $\mathcal{O}(h^2+kh)$ and $\mathcal{O}(4h^2)$\cite{graves2016hybrid}, respectively ($k$ is item embedding size, and $h$ is hidden state size). Based on the time complexity of the controller and EMN, the total time consumption of MNR and CmnRec is $(5h^2+kh)*T$ and $(h^2+kh)*T+4mh^2$, respectively. When $h=2k$, the time consumption ratio of MNR and CmnRec is $1\leq \frac{8}{3+8\frac{m}{T}}\leq \frac{8}{3}$. In practice, a larger acceleration can be obtained due to there are many other complex operations when EMN processes memory operations, such as memory addressing operations~\cite{graves2016hybrid}.

\section{\label{find_upper_bound}Find upper bound}
The update equation of standard RNN hidden state is:
\begin{equation}
\boldsymbol{h}_t=tanh(\boldsymbol{U}\boldsymbol{h}_{t-1}+\boldsymbol{W}\boldsymbol{v}_{t}+\boldsymbol{b})  \label{eq:s_rnn} 
\end{equation}
Starting from the derivative of $\boldsymbol{v}_t$, we make use of the result $\vartheta _q=B\left \| \boldsymbol{U} \right \|$ which is shown in~\cite{le2019learning}, where $B$ is the bound of \\ $\left \| diag(tanh^{'}(\boldsymbol{U}\boldsymbol{h}_{t-1}+\boldsymbol{W}\boldsymbol{v}_{t}+\boldsymbol{b})) \right \|$. 
Then we take the derivative of $\boldsymbol{h}_{t-1}$. 
In Eq.(~\ref{eq:s_rnn}), $\boldsymbol{U}\boldsymbol{h}_{t-1}$ and $\boldsymbol{W}\boldsymbol{v}_{t}$ are interchangeable. 
Referring to the solution of $\vartheta _q$, we replace $\boldsymbol{W}$ with $\boldsymbol{U}$, and $\boldsymbol{v}_t$ with $\boldsymbol{h}_{t-1}$. As a result, we  have $\vartheta _p=B\left \| \boldsymbol{W} \right \|$.
\\\\For LSTM, the update equation is:
\begin{align}
\boldsymbol{c}_t & =\sigma (\boldsymbol{U}_{f}\boldsymbol{h}_{t-1}+\boldsymbol{W}_{f}\boldsymbol{v}_{t}+\boldsymbol{b}_{f}) \odot \boldsymbol{c}_{t-1}   \nonumber \\
& + \sigma (\boldsymbol{U}_{i}\boldsymbol{h}_{t-1}+\boldsymbol{W}_{i}\boldsymbol{v}_{t}+\boldsymbol{b}_{i}) \odot 
tanh(\boldsymbol{U}_{z}\boldsymbol{h}_{t-1}+\boldsymbol{W}_{z}\boldsymbol{v}_{t}+\boldsymbol{b}_{z}) \nonumber\\
\boldsymbol{h}_t & =tanh(\boldsymbol{U}_{o}\boldsymbol{h}_{t-1}+\boldsymbol{W}_{o}\boldsymbol{v}_{t}+\boldsymbol{b}_{o}) \odot  tanh(\boldsymbol{c}_{t}) \label{eq:lstm}
\end{align}
Following the same way, we start from the derivative of $\boldsymbol{v}_t$. According to the results in~\cite{le2019learning}, $\vartheta _q >0$ is equivalent to $\vartheta _q q_{i,t}\geq q_{i-1,t}$. To
solve the problem of $\vartheta _p$,  we  define:
\begin{equation}
\Phi _{\varrho} (\boldsymbol{h}_{t-1},\boldsymbol{v}_{t},\boldsymbol{U}_{\varrho},\boldsymbol{W}_{\varrho},\boldsymbol{b}_{\varrho})=
\left\{
\begin{array}{lr}
\boldsymbol{U}_{f}\boldsymbol{h}_{t-1}+\boldsymbol{W}_{f}\boldsymbol{v}_{t}+\boldsymbol{b}_{f},&\\
\boldsymbol{U}_{i}\boldsymbol{h}_{t-1}+\boldsymbol{W}_{i}\boldsymbol{v}_{t}+\boldsymbol{b}_{i},&\\
\boldsymbol{U}_{z}\boldsymbol{h}_{t-1}+\boldsymbol{W}_{z}\boldsymbol{v}_{t}+\boldsymbol{b}_{z},&\\
\boldsymbol{U}_{o}\boldsymbol{h}_{t-1}+\boldsymbol{W}_{o}\boldsymbol{v}_{t}+\boldsymbol{b}_{o}.& \\ 
\end{array}
\right. \nonumber
\end{equation}
where $\varrho \in \{f,i,o,z\} $. 
In Eq.(~\ref{eq:lstm}), we swap $\boldsymbol{U}_{\varrho}\boldsymbol{h}_{t-1}$ and $\boldsymbol{W}_{\varrho}\boldsymbol{v}_{t}$. As a result, $\boldsymbol{h}_t$ and $\boldsymbol{c}_t$ remain unchanged, which means $\boldsymbol{h}_t$, $\boldsymbol{U}_{\varrho}\boldsymbol{h}_{t-1}$ and $\boldsymbol{W}_{\varrho}\boldsymbol{v}_{t}$ are equivalent. 
Following the solution of $\vartheta _q$ in standard RNN, we replace $\boldsymbol{W}_{\varrho}$ with $\boldsymbol{U}_{\varrho}$, and $\boldsymbol{v}_t$ with $\boldsymbol{h}_{t-1}$. 
As a result, we have $\vartheta _p >0$ to guarantee $\vartheta _p p_{i,t}\geq p_{i-1,t}$.

\section{\label{find_depen}The contribution of memory slots}
Accroding to Eq.(\ref{rnn_con}) , the contribution of each slot can be measured as $\sum_{i=0}^{l_r -1} \vartheta _q^{i} q_{t_r,t_r},(r=\{1,2,3...m\}) $, i.e.,
\begin{align}
q_{t_{r},t_{m}} &= \left \| \frac{\partial \boldsymbol{h}_{t_m} }{\partial  \boldsymbol{v}_{t_r}} \right \| =  \left \| \frac{\partial \boldsymbol{h}_{t_m} }{\partial  \boldsymbol{h}_{t_r}} \frac{\partial \boldsymbol{h}_{t_r} }{\partial  \boldsymbol{v}_{t_r}} \right \|  \nonumber   \\ 
& = p_{t_r,t_m}q_{t_r,t_r} \leq \vartheta _p^{t_m-t_r}  p_{t_m,t_m}q_{t_r,t_r} \nonumber\\
& = \vartheta _p^{T-t_r}  p_{T,T}q_{t_r,t_r}   \qquad (t_m=T)    \label{eq:rela}
\end{align}
For the sake of analysis,  assuming the contribution of $\boldsymbol{v}$ and $\boldsymbol{h}_t$ at time step $t$ is constant, i.e., $q_{t_1,t_1}=q_{t_2,t_2}=...=q_{t_m,t_m}$. Then we can get $\sum_{i=0}^{l_r -1} \vartheta _q^{i} q_{t_r,t_r}= \sum_{i=0}^{l_m -1} \vartheta _q^{i}\vartheta _p^{T-t_m} p_{T,T}q_{t_r,t_r} $. Considering that $p_{T,T}q_{t_r,t_r}$ is same for every $r$, $\sum_{i=0}^{l_m -1} \vartheta _q^{i}\vartheta _p^{T-t_m} p_{T,T}q_{t_r,t_r} $ can be simplified to $\sum_{i=0}^{l_m -1} \vartheta _q^{i}\vartheta _p^{T-t_m} $.

\ifCLASSOPTIONcompsoc
  \section*{Acknowledgments}
\else
\fi
This work is partially supported by the National Natural Science Foundation of China under Grant (No. 62032013, 61972078, 61602197, L1924068), Hebei Natural Science Foundation No.G2021203010 and CCF-AFSG Research Fund under Grant No.RF20210005.




\bibliographystyle{IEEEtran}
\bibliography{sample-base}
%
\begin{IEEEbiography}[{\includegraphics[width=1in,height=1.25in,clip,keepaspectratio]{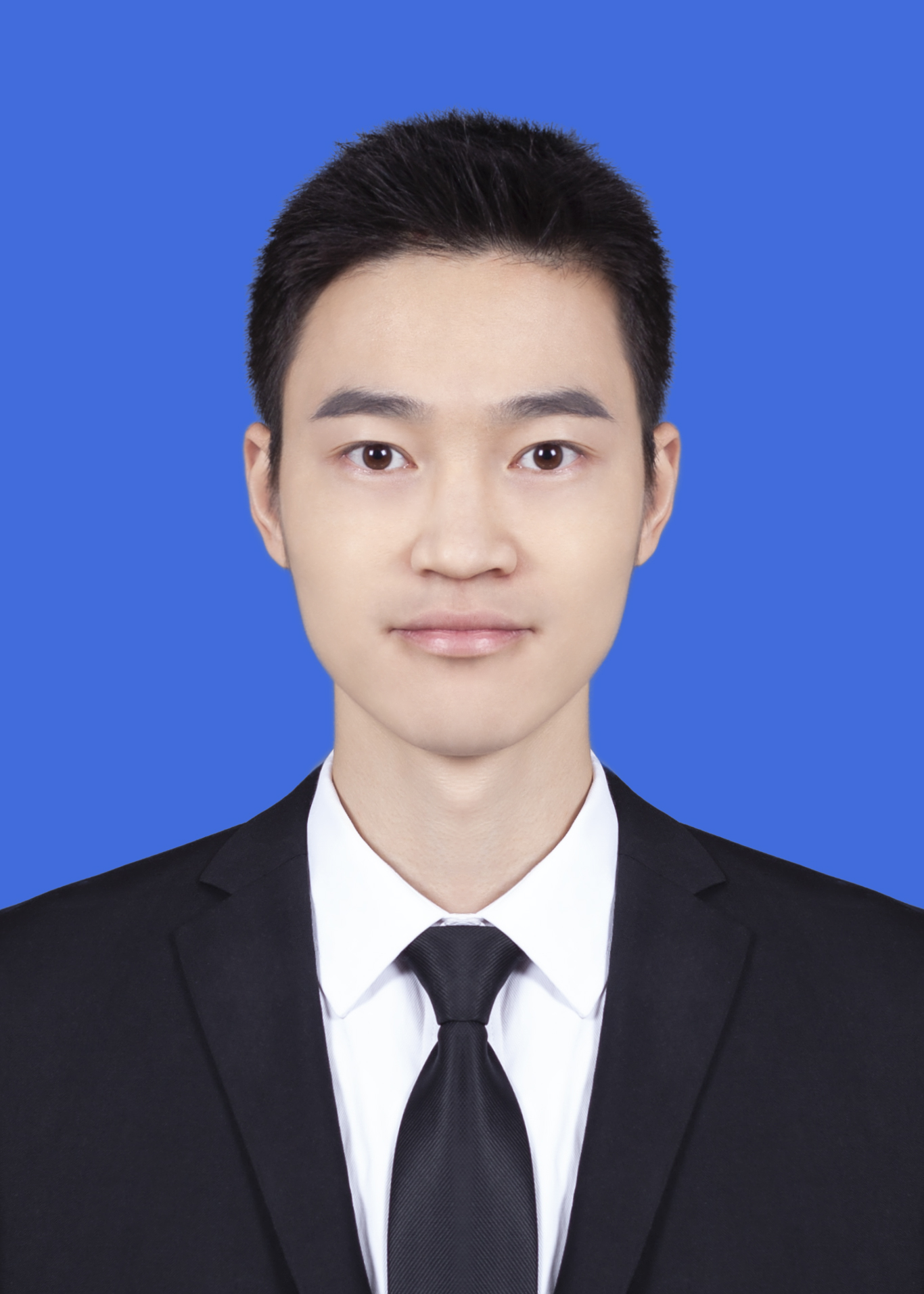}}]{Shilin Qu} received the BS and MS degrees in software engineering from Northeastern University, China, in 2017 and 2020 respectively. He is currently working toward a PhD degree in the Faculty of Information Technology, Monash University, Australia. His research interests are in the area of graph representation learning and recommender system.
\end{IEEEbiography}
\begin{IEEEbiography}[{\includegraphics[width=1in,height=1.25in,clip,keepaspectratio]{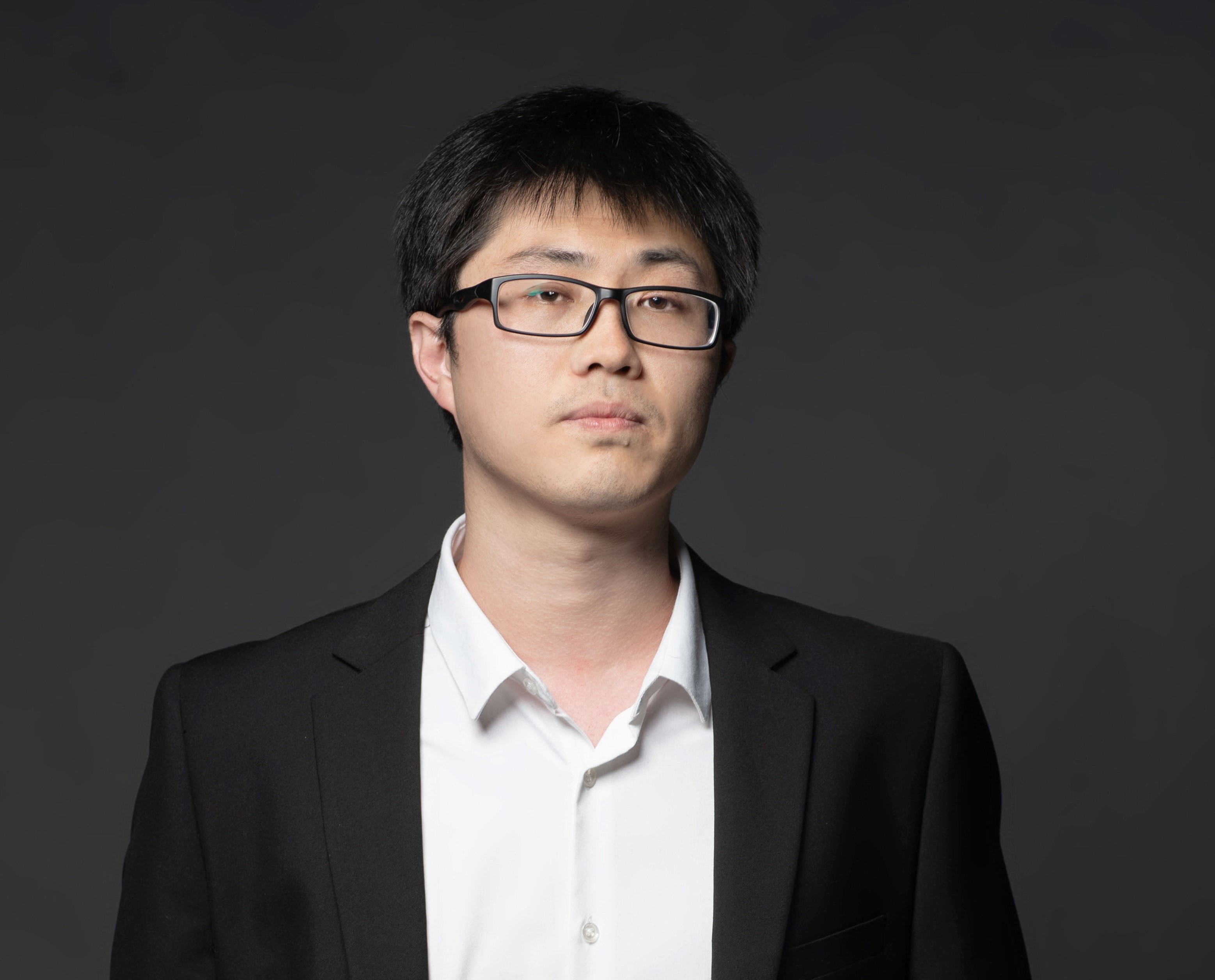}}]{Fajie Yuan} is currently an assistant professor at Westlake University. Prior to that, he was a senior AI researcher at Tencent working on recommender systems and machine learning. He received his Ph.D. degree from the University of Glasgow. His main research interests include deep learning and transfer learning and their applications in recommender systems.
\end{IEEEbiography}
\begin{IEEEbiography}[{\includegraphics[width=1in,height=1.25in,clip,keepaspectratio]{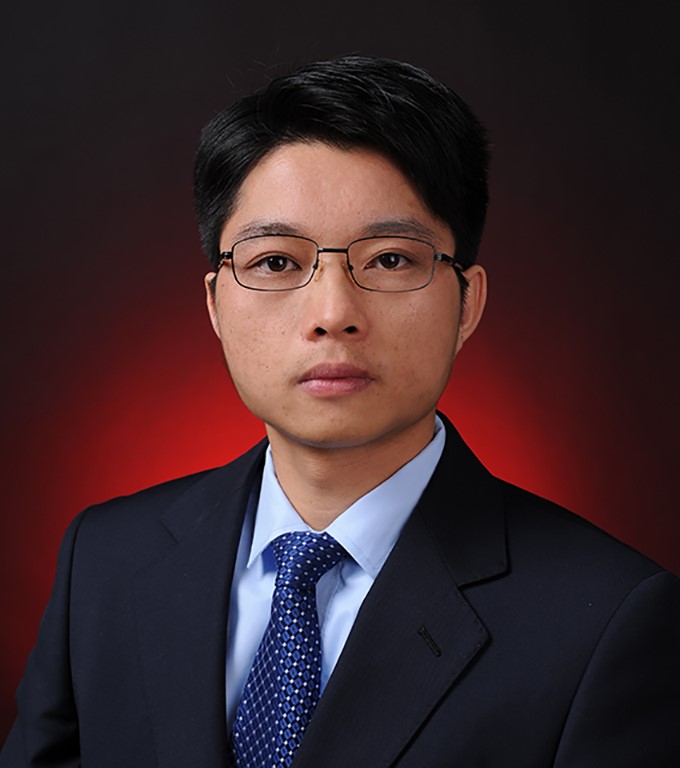}}]{Guibing Guo} is currently a Professor with the Software College, Northeastern University, Shenyang, China. He received the Ph.D. degree in computer science from Nanyang Technological University, Singapore, in 2015. His research interests include recommender systems, deep learning, natural language processing, and data mining.
\end{IEEEbiography}

\begin{IEEEbiography}[{\includegraphics[width=1in,height=1.25in,clip,keepaspectratio]{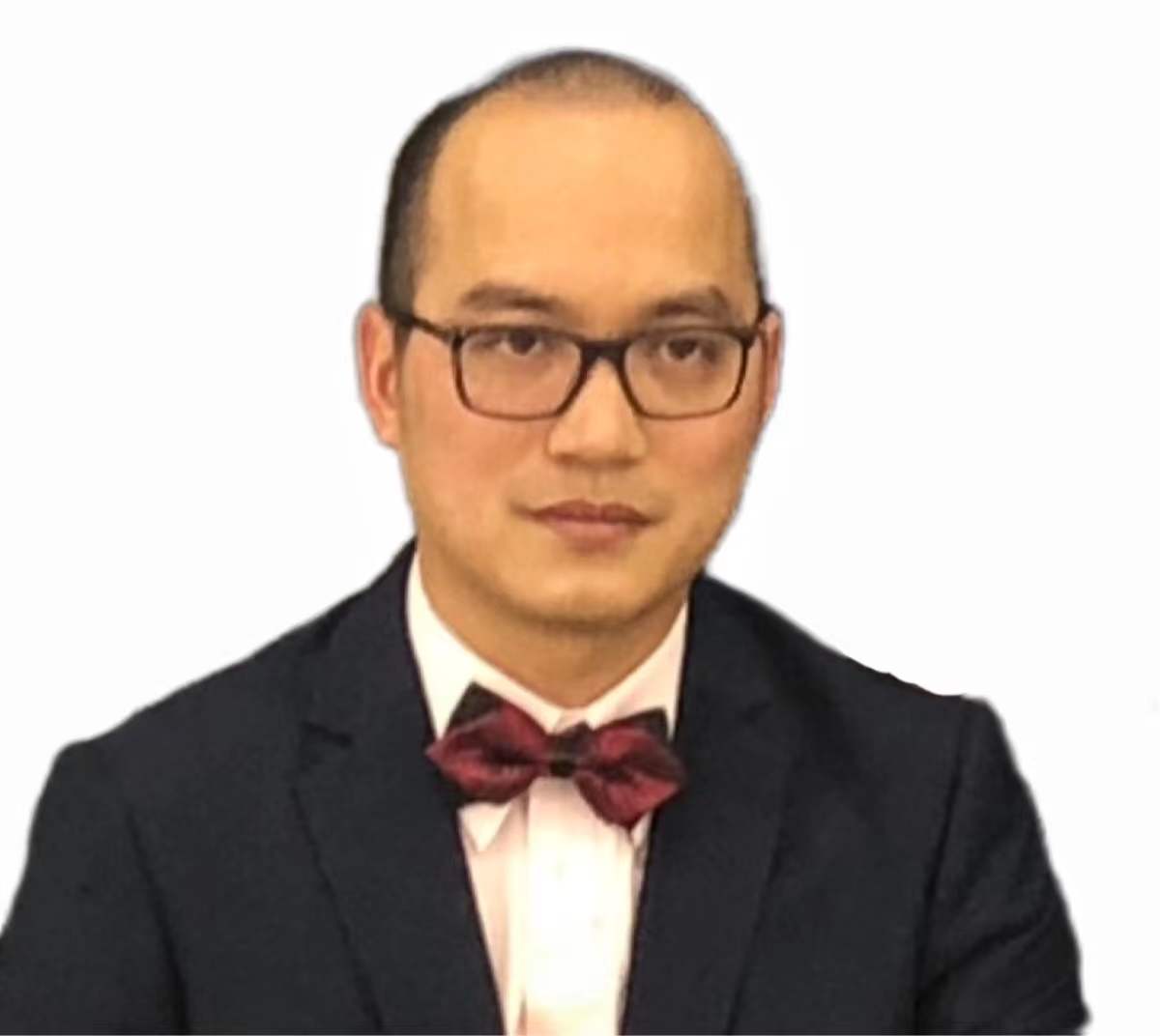}}]{Liguang Zhang} is a technician at Tencent. His mainly responsibility is business recommendation of KANDIAN.
\end{IEEEbiography}

\begin{IEEEbiography}[{\includegraphics[width=1in,height=1.25in,clip,keepaspectratio]{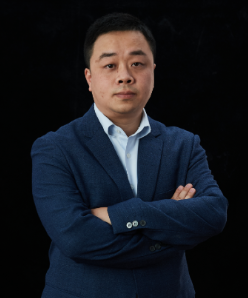}}]{Wei Wei}  received the Ph.D. degree from the Huazhong University of Science and Technology, China, in 2012. He is currently an Associate Professor with School of Computer Science and Technology and the Director of Cognitive Computing and Intelligent Information Processing (CCIIP) Laboratory in Huazhong University of Science and Technology, China. His major research interests include information retrieval, natural language processing, artificial intelligence, data mining (text mining), statistics machine learning, social media analysis and mining recommender system.
\end{IEEEbiography}

\end{document}